# Relativity and quantum mechanics: Jorgensen[1] revisited
## 1. Introduction

Bernhard Rothenstein, "Politehnica" University of Timisoara, Physics Department, Timisoara, Romania.  brothenstein@gmail.com

*Abstract. We first define the functions which ensure the transformation of momentum and energy of a tardyon, the transformation of the wave vector and the frequency of the associated wave. Having done this, we show that they ensure the relativistic invariance of the quotient between momentum and wave vector and between energy and frequency if the product between particle velocity u and phase velocity w is a relativistic invariant (uw=$c^2$), a condition which is a natural combination of special relativity theory and quantum mechanics.*

Jorgensen[1] starts his paper entitled "Relativity and quantum" by stating that beams of entities, such as electrons, may produce diffraction patterns. These patterns may be interpreted in terms of particles and waves. One obvious question concerning this phenomenon is **what is the functional relation between the momentum of the entity and its momentum?** While this relation is well known, it is of interest to look for another way to arrive at this function by using special relativity theory and the fundamental observation that the mathematical form of a law of nature cannot contain any parameters relating to more than one reference frame.

The purpose of our paper is to analyze the properties involved in the transformation of the space-time coordinates of the same event, of the mass (energy) and momentum of the same tardyon and of the frequency and the wave vector of a wave propagating at subluminal velocities; this last enables us to introduce the concept of phase velocity.

## 2. What do transformation equations share in common?

Consider an electron that moves with velocity u relative to the inertial reference frame I along a direction that makes an angle θ with the positive direction of the OX axis. After a given time of propagation t the electron arrives at a point $M(x = r\cos\theta = ut\cos\theta; y = r\sin\theta = ut\sin\theta)$ generating the event $E(x = r\cos\theta = ut\cos\theta; y = r\sin\theta = ut\sin\theta; t = r/u = \sqrt{x^2 + y^2}/u)$, using Cartesian (x,y) and polar (r,θ) coordinates in a two space dimensions approach. Detected from the I' inertial reference frame that moves with constant velocity V relative to I in the positive direction of the overlapped OX(O'X') axes, the same event is $E'(x' = r'\cos\theta' = u't'\cos\theta', t' = r'/u' = \sqrt{x'^2 + y'^2}/u')$. In accordance with the Lorentz transformations for the space-time coordinates of the same event, the space coordinates become

$$r' = \gamma r \left[(\cos\theta - V/u)^2 + (1-V^2/c^2)\sin^2\theta\right]^{1/2} = \gamma r \Phi_1(u,V,c,\theta) \quad (1)$$

$$t' = \gamma t \left(1 - \frac{Vu}{c^2}\cos\theta\right) = \gamma t \Phi_2(u,V,\theta,c). \quad (2)$$

We underline an important property of the functions defined above: both of them have the same limit for u→c, i.e.

$$\Phi_{1,c} = \lim_{u \to c} \Phi_1(u,V,\theta,c) = 1 - \frac{V}{c}\cos\theta \quad (3)$$

and

$$\Phi_{2,c} = \lim_{x \to \infty} \Phi_2(U,V,\theta,c) = 1 - \frac{V}{c}\cos\theta \quad (4)$$

### 2.2. Transformation equations for the mass, energy and momentum of the same electron

Let m, E and **p** be the mass, energy and momentum of an electron when detected from I and m', E' and **p'** when detected from I'. By definition

$$p = mu \quad (5)$$

$$p' = m'u' \quad (6)$$

as long as we are interested in the magnitudes of the physical quantities involved. Combining (5) and (6), we obtain

$$\frac{p'}{m'} = \frac{p}{m}\frac{u'}{u} = \frac{p}{m}\frac{\Phi_1(u,V,\theta,c)}{\Phi_2(U,V,\theta,c)}. \quad (7)$$

Equation (7) suggests considering that

$$p' = F(V)\Phi_1(u,V,\theta,c) \quad (8)$$

and

$$m' = F(V)\Phi_2(u,V,\theta,c) \quad (9)$$

where F(V) represents an unknown function which depends only on the relative velocity of the inertial reference frames involved. We obtain this function by imposing u=0 in (9). Under such conditions, observers from I measure the rest mass of the electron as $m_0$, whereas observers from I' measure its inertial mass m given by

$$m = F(V)m_0 \Phi_{2,u=0} = F(V)m_0. \quad (10)$$

Experiment[2] and special relativity combined with conservation laws of momentum and energy[3] leads to the following relationship between rest mass and inertial mass of the same electron

$$m = \gamma m_0 \quad (11)$$

with the result that

$$F(V) = \gamma = \frac{1}{\sqrt{1-\frac{V^2}{c^2}}}. \quad (12)$$

Equations (8) and (9) become
$$p' = \gamma p\Phi_1(u,V,\theta,c) \tag{13}$$
$$m' = \gamma m\Phi_2(u,V,\theta,c). \tag{14}$$
Multiplying both sides with $c^2$, we obtain that energy transforms as
$$E' = \gamma E\Phi_2(u,V,\theta,c). \tag{15}$$

Jorgensen[1] proposes a transparent criterion for finding out if a combination of physical quantities is a relativistic invariant or not. Mimicking his approach we start, with
$$\frac{p'}{m'} = \frac{p}{m}\frac{1-V/u}{1-uV/c^2} = \frac{p}{m}\left[1+\left(\frac{u}{c^2}-\frac{1}{u}\right)\left(V+\frac{u}{c^2}V^2+...\right)\right] \tag{16}$$
in the case of a single space dimension ($\theta=\theta'=0$). We see that $\frac{p}{m} = \frac{p'}{m'}$ only in the case of a photon (u=u'=c), in which case the quotient between the momentum and the inertial mass of the photon is the invariant c.

### 2.3. Transformation equation for frequency and wave vector

Moller[3] derives transformation equations for the parameters introduced in order to characterize a plane wave propagating with subluminal velocity. The parameters mentioned are unit vector **n** that characterizes the direction in which the wave propagates, the phase velocity w, the frequency f, the wave vector $\mathbf{k} = \frac{f}{w}\mathbf{n}$, and the wavelength $\lambda$ when detected from I, but **n'**,w',f', $\mathbf{k'} = \frac{f'}{w'}\mathbf{n'}$ and $\lambda'$ when detected from I'. Using results obtained by this author and extending them to the case of the wave vector, we have the following transformation equations
$$f' = \gamma f\left(1-\frac{V}{w}\cos\theta\right) = \gamma\Psi_1(V,w,\theta) \tag{17}$$
for the frequencies of the oscillations taking place in the wave and
$$k' = \gamma k\sqrt{1-\frac{2Vw}{c^2}\cos\theta+\frac{V^2}{c^2}\left(\frac{w^2}{c^2}-\sin^2\theta\right)} = \gamma k\Psi_2(u,V,\theta,c) \tag{18}$$
for the magnitude of the wave vector. The functions $\psi_1$ and $\psi_2$ defined above have the same limit for w=c equal to
$$\Psi_c = \lim_{x\to\infty}\left(1-\frac{V}{w}\cos\theta\right) = \lim_{x\to\infty}\sqrt{1-\frac{2Vw}{c^2}\cos\theta+\frac{V^2}{c^2}\left(\frac{w^2}{c^2}-\sin^2\theta\right)} = 1-\frac{V}{c}\cos\theta \tag{19}$$

### 3. Quantum mechanics and special relativity

De Broglie[4] associates a wave that propagates with phase velocity w to a tardyon moving with velocity u, establishing the following relationships between momentum, wave vector, energy and frequency

$$\frac{p}{k} = h \tag{20}$$

$$\frac{E}{\nu} = h \tag{21}$$

where $\nu$ and h stand for the frequency of the oscillations in the wave and for Planck's constant, respectively. Because universal constants are relativistic invariants, we expect that the combinations of physical quantities P/k and E/$\nu$ should have the same magnitude in all inertial reference frames in relative motion. As we can see, De Broglie's equations (20) and (21) combine a physical quantity which characterizes the particle property (P,E) and a physical quantity that characterizes the wave property (k,$\nu$). Taking into account the results obtained above, we can present (20) and (21) as

$$\frac{p'}{k'} = \frac{p}{k} \frac{\Phi_1(u,V,\theta,c)}{\Psi_2(V,w,\theta,c)} = \frac{p}{k} \frac{\sqrt{\left(\cos\theta - \frac{V}{u}\right)^2 + \left(1 - \frac{V^2}{c^2}\right)\sin^2\theta}}{\sqrt{1 - \frac{2Vw}{c^2}\cos\theta + \frac{V^2}{c^2}\left(\frac{w^2}{c^2} - \sin^2\theta\right)}} \tag{22}$$

$$\frac{E'}{\nu'} = \frac{E}{\nu} \frac{\Phi_2(u,V,\theta,c)}{\Psi_1(V,w,\theta)} = \frac{1 - \frac{Vu}{c^2}\cos\theta}{1 - \frac{V}{w}\cos\theta} \tag{23}$$

Analyzing the properties of the functions involved in the transformation process, we see that all of them have the same limit, not only for u=w=c, but also for

$$uw = c^2. \tag{24}$$

Moller[3] derives condition (24) by imposing the condition that the particle moves and the wave propagates along the same direction. A combination of quantum mechanics and special relativity leads to the same result and goes as follows. Consider that, in accordance with De Broglie, we have

$$k = \frac{m_0 u}{h\sqrt{1 - \frac{u^2}{c^2}}} \tag{25}$$

and using the relativistic energy

$$E = \frac{m_0 c^2}{\sqrt{1 - \frac{u^2}{c^2}}} = h\nu \tag{26}$$

where $m_0$ stands for the rest mass of the particle, combining (25) and (26), we obtain
$$uw = c^2 \tag{27}$$
i.e. the condition which ensures the invariance of p/k and of E/$\nu$, and demonstrates the intimate relationship between special relativity theory and quantum mechanics.

**Acknowledgment**

I wish to acknowledge the help of A. Pearlstein in the elaboration of the present paper who guided me through Jorgensen's paper.